\documentclass[aps,pra,preprint,superscriptaddress]{revtex4}

\usepackage[usenames,dvipsnames]{color}
\usepackage{amssymb}
\usepackage{graphicx}
\usepackage{cancel}
\usepackage{amsmath}

\begin{document}

\title{Metastable Bose-Einstein Condensation in a Strongly Correlated Optical Lattice}
\author{David McKay}
\affiliation{James Franck Institute, University of Chicago, 929 E 57th St, Chicago, IL 60637}
\author{Ushnish Ray}
\affiliation{Department of Physics, University of Illinois, 1110 W Green St, Urbana, IL 61801}
\author{Stefan Natu}
\affiliation{Joint Quantum Institute, University of Maryland, College Park, MD 20742}
\author{Philip Russ}
\affiliation{Department of Physics, University of Illinois, 1110 W Green St, Urbana, IL 61801}
\author{David Ceperley}
\affiliation{Department of Physics, University of Illinois, 1110 W Green St, Urbana, IL 61801}
\author{Brian DeMarco}
\affiliation{Department of Physics, University of Illinois, 1110 W Green St, Urbana, IL 61801}
\date{\today}

\begin{abstract}
We experimentally and theoretically study the peak fraction of a Bose-Einstein condensate loaded into a cubic optical lattice as the lattice potential depth and entropy per particle are varied. This system is well-described by the superfluid regime of the Bose-Hubbard model, which allows for comparison with  mean-field theories and exact quantum Monte Carlo (QMC) simulations. Despite correcting for systematic discrepancies between condensate fraction and peak fraction, we discover that the experiment consistently shows the presence of a condensate at temperatures higher than the critical temperature predicted by QMC simulations. This metastability suggests that turning on the lattice potential is non-adiabatic. To confirm this  behavior, we compute the timescales for relaxation in this system, and find that equilibration times are comparable with the known heating rates. The similarity of these timescales implies that turning on the lattice potential adiabatically may be impossible.  Our results point to the urgent need for a better theoretical and experimental understanding of the timescales for relaxation and adiabaticity in strongly interacting quantum gases, and the importance of model-independent probes of thermometry in optical lattices.
\end{abstract}

\maketitle
\section{Introduction}

Research efforts using ultracold quantum gases trapped in optical lattices to study models of strongly correlated matter, such as cuprate superconductors, are at a crossroads.  Rapid experimental progress has enabled emulation of ever more interesting and complex Hamiltonians, including those that go beyond conventional condensed matter paradigms. However, in most cases, the ultra-low temperatures required to study exotic many-particle ground states have been inaccessible to experiments \cite{mckay:2011}.  For example, experiments have prepared a gas at temperatures close to the N\'{e}el temperature for a fermionic atoms trapped in a cubic lattice \cite{hart:2014}, which realize the Hubbard model.  However, there are no known viable methods for reaching the orders-of-magnitude lower temperature required to achieve the putative $d$-wave superfluid state.

The current approach to achieving low temperatures in experiments to achieving low temperatures is to cool a gas confined in a parabolic trap and then slowly turn on an optical lattice potential. Reaching lower temperatures in the lattice requires achieving lower entropy per particle in the trap.  At best turning on the lattice preserves entropy per particle.  Generally, imperfections in the experiment, such as spontaneous scattering of the lattice light, and non-adiabaticity will lead to heating, and the entropy per particle of the gas in the lattice will be higher than the initial state.  There are also proposals to cool the relevant degrees of freedom for atoms in the lattice, which have not been realized in experiments.  In either approach, understanding and verifying equilibration and measuring temperature are critical to exploring low entropy states.

Understanding relaxation and equilibration dynamics in strongly correlated systems is especially challenging.  Exact dynamics can only be calculated reliably for one-dimensional systems and for short times.  In two and three dimensions, we are often forced to rely on approximate time-dependent mean-field theories, which ignore correlation effects, or exact simulations of small numbers of particles, which may not reproduce experiments.  Ultracold gas experiments introduce an additional complication in that they are isolated from the environment and a thermal bath is absent.  Equilibration and thermalization in closed quantum systems via interparticle interactions is an open topic rife with unresolved questions \cite{PhysRevA.43.2046,rigol2008thermalization}.

Measuring temperature in closed quantum systems is also challenging.  In systems connected to a reservoir, such as electronic solids, thermal contact is maintained with a carefully calibrated thermometer.  In contrast, in closed systems, temperature must be measured via its relation to an observable derived from the system itself.  Such a procedure can be problematic when the physics and energy spectrum of the system of interest are not fully understood.

In this paper, we use a combination of experiment, QMC simulations, and semi-analytic theory to explore equilibration and temperature in a prototypical strongly correlated system: the Bose-Hubbard model realized using ultracold bosonic $^{87}$Rb atoms trapped in an optical lattice \cite{greiner:2002,jaksch:1998}.  In the BH model, particles tunnel between adjacent lattice sites with energy $t$, and particles interact on the same site with energy $U$.  Changing the ratio of $t/U$ realizes a quantum phase transition between superfluid and Mott insulating states \cite{PhysRevB.40.546,sachdev}.  An important feature of the BH model is that the equilibrium properties can be computed using a variety of theoretical tools, and thus the experiments and theory can be benchmarked against one another.  If $t/U<1$, strong coupling expansions \cite{PhysRevB.53.2691,sengupta:2005}, Gutzwiller theory \cite{sheshadri:1993}, and dynamical mean field theory \cite{anders:2010} can be used. Conversely, if $t/U>1$, the Hartree-Fock-Bogliubov-Popov (HFBP) approximation can be employed \cite{lin:2008,rey:2003,oosten:2001}.  In this paper, we comprehensively test the ability of these approximate methods to predict condensate fraction, which is a primary experimental observable, by comparing to quantum Monte Carlo simulations.  Quantum Monte Carlo (QMC) \cite{kashurnikov:2002,wessel:2004} is statistically exact for the BH model for all ranges of $t/U$, since the sign problem is absent.  For all theoretical and numerical predictions, we include the effects of the parabolic trapping potential that is present in combination with the lattice in experiments.

To experimentally test equilibration, we compare peak fraction $f_0$ \cite{schori:2004,spielman:2008,lin:2008,trotzky:2010} measured for lattice gases prepared at varied temperature and $t/U$ (in the superfluid regime) to predictions from QMC simulations.  Gases are cooled to different temperatures in the trap before the lattice is turned on.  Temperature and entropy per particle are straightforwardly determined for the weakly interacting trapped gas using time-of-flight (TOF) imaging and a semi-ideal model \cite{PhysRevA.58.2423}.  After the lattice is turned on slowly compared with the tunneling and interaction timescales, peak fraction---which we show using QMC simulations is closely related to condensate fraction $n_0$ and can be used as an accurate thermometer---is measured via TOF imaging.  If the lattice turn-on is isentropic and the gas remains in thermal equilibrium, then the measured peak fraction should match the QMC prediction for the entropy per particle measured before the lattice is turned on.  At fixed entropy per particle, the peak and condensate fraction drops at higher lattice potential depths because of strong inter-particle interactions.  Generally, the peak and condensate fraction will be lower (and temperature higher) in the lattice than predicted for the adiabatic case because of heating processes.

At low entropy per particle, we observe good agreement with QMC predictions for peak fraction.  However, at higher temperatures, $f_0$ is larger than predicted, implying lower entropy per particle in the lattice than the initial state, in apparent contradiction with the second law of thermodynamics.  Above the critical entropy per particle for superfluidity in the lattice, a regime that has not been widely explored in experiments, we find that $f_0$ is non-zero.  When such disagreement is present, we show that there is no clear adiabatic timescale in the experiment by measuring $f_0$ for varied lattice-turn-on times.

While a full modeling of the dynamics present in the experiment is beyond the scope of this paper, we obtain a rough estimate for the timescales for equilibration and the lifetime of the condensate by calculating the Landau damping time in the lattice.  We find that the timescale for thermalization can be much longer than the tunneling and interaction timescales at high entropy per particle.  Furthermore, the condensate lifetime is comparable to the heating timescale, making discrimination of adiabaticity in the experiment difficult.  Our measurements in combination with these calculations imply the disagreement between QMC and the experiment is likely due to a lack of equilibrium and metastability of the condensate in the lattice, making thermal equilibrium difficult to achieve and challenging to distinguish from heating.

This paper is organized as follows.  In Section II, we discuss the experimental procedure along with the measurement protocols used to obtain $f_0$ as well as the entropy per particle $S/N$.  In Section III, we describe how the equilibrium $n_0$ and $f_0$ are computed using full-scale QMC {\it ab initio} simulations and mean-field theories.  We show that the mean-field theory results are generally in poor agreement with exact QMC predictions.  In Section IV, we discuss our results and show a comparison between the experimentally measured $f_0$ and QMC predictions.   The discrepancy with experimental measures of $f_0$ inevitably points to lack of adiabaticity during loading and the need for a better understanding of the timescales of dynamics in this system. Subsequently in Section V, we attempt to obtain a rough estimate of the timescales for relaxation of the peak fraction by calculating the Landau damping time in the lattice for the experimentally measured entropies. Our calculations suggest that the timescale for equilibration is comparable to the timescale for heating, implying that the lack of agreement between the theoretical and experimental results may be due to a non-equilibrium effect.  In Section VI, we present concluding remarks and discuss future research directions.

\section{Experiment}

The details of our $^{87}$Rb BEC lattice apparatus are described in Ref. \cite{mckay:2009}. We prepare approximately 200,000 $^{87}$Rb $\left|F=1,m_F=-1\right\rangle$ atoms confined in a harmonic trap that is formed from a single-beam optical dipole trap at 1064~nm traveling perpendicular to gravity and a magnetic quadrupole trap balancing gravity and providing additional harmonic confinement in the two horizontal directions. We control the entropy per particle by evaporating to different depths of the dipole trap, after which we ramp the depth to a constant value with a mean harmonic trap frequency of $\omega_0=2\pi\left(35.78\pm6\right)$~Hz. To keep the atom number roughly fixed for the different temperatures, we selectively adjust the efficiency of the evaporation.

A cubic optical lattice is formed by ramping on three sets of retro-reflected lattice beams at $\lambda=812$~nm over $100$ ms (for most of the data in this paper). The lattice potential depth is characterized as $sE_R$, where $E_R=(h/\lambda)^2/2m=2.31\times10^{-30}$~J is the recoil energy. We use Kaptiza-Dirac diffraction to calibrate $s$ to within 7\%. The lattice beams add additional harmonic confinement, and the overall confinement is well-described by $\omega=\sqrt{\omega_0^2+\frac{8sE_R}{m(2\pi)^2(120\mu m)^2}}$. After turning on the lattice and waiting for 10~ms, we turn off all potentials and allow the atoms expand for $20$~ms time-of-flight. We optically pump atoms into the $F=2$ state, and absorption image on the $F=2$ to $F^{'}=3$ cycling transition.

The optical depth ($OD$) in absorption images saturates for $OD>3$  (i.e., high column density) because of non-absorbable light in the imaging laser beam.  This is a complication for imaging non-zero temperature but condensed gases, which possess high-density peaks and a low-density thermal component.  To increase the dynamic range of imaging and obtain high signal-to-noise ratio data for the condensate and non-condensate atoms, we take a series of two images in separate measurements. We produce high $OD$ images by transferring all the atoms into $F=2$, and low $OD$ images are acquired by pumping a fraction of the atoms so that the $OD<2$.  High $OD$ images are used to measure the atoms outside of the condensate peaks, and low $OD$ images are employed to measure the atoms in the peaks (which are  saturated in the high $OD$ images).

\begin{figure}[h]
\includegraphics[width=0.9\textwidth]{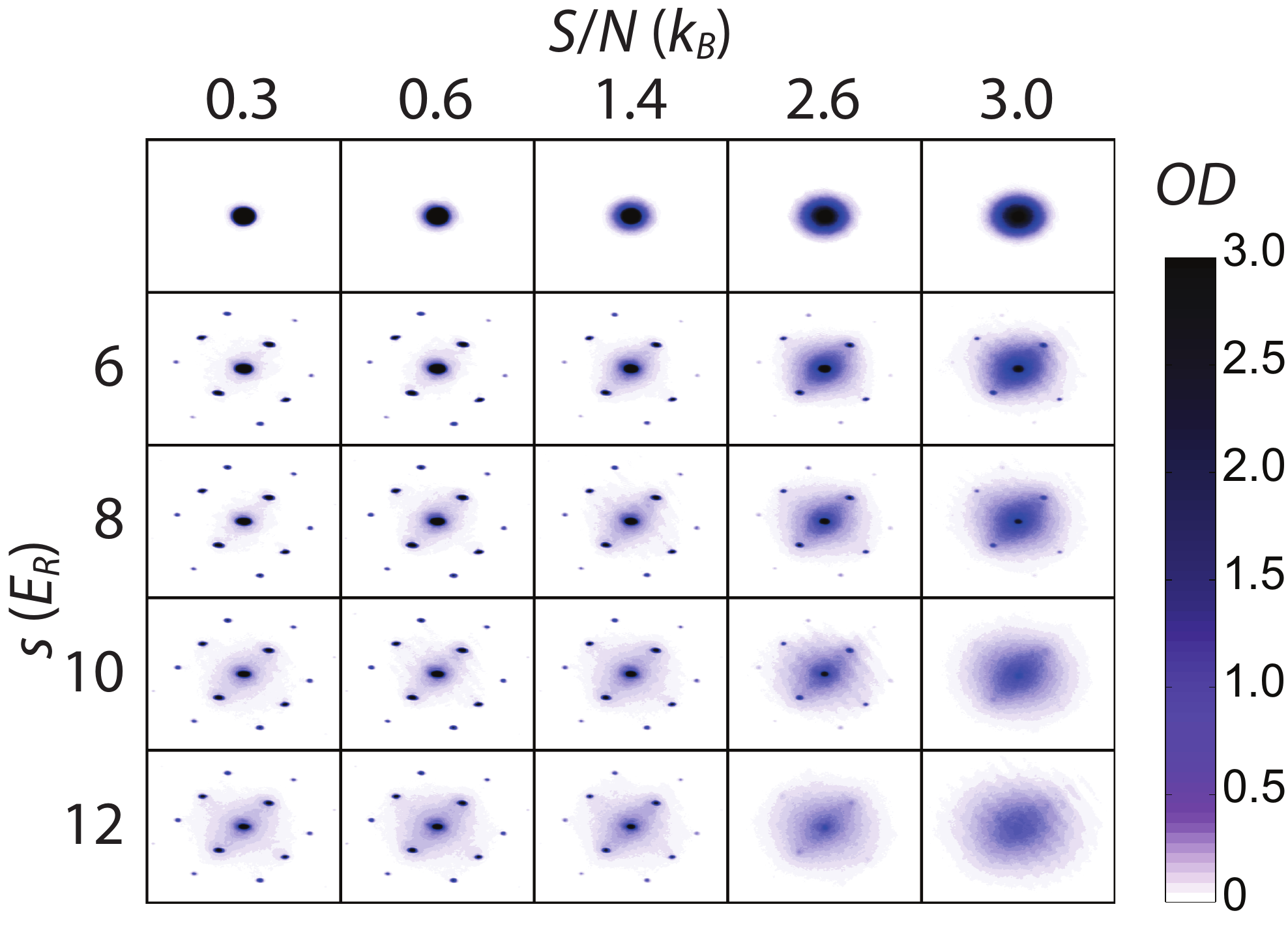}
\caption{(Color Online) High $OD$ TOF images for a variety of lattice depths (in units of $E_\text{R}$) and initial entropies averaged over 7 iterations. The top line shows the gas released from the harmonic trap; the starting conditions were almost identical for all lattice depths.  The $OD$ saturates at approximately 3 because of non-absorbable light in the laser beam used for imaging. \label{fig:1}}
\end{figure}

A typical experimental sequence at a given lattice depth and entropy per particle involved 7 sets of 4 images: a high $OD$ and low $OD$ image in the harmonic trap, and a high $OD$ and low $OD$ image in the lattice. All quantities presented in this paper are averages over these 7 sets of images. Figure \ref{fig:1} illustrates the resulting averaged high $OD$ images.

To determine $S/N$ for the gas before turning on the lattice, images for gases released from the harmonic trap were fit using a multi-step scheme similar to that employed in Ref.~\cite{szczepkowski:2009} to determine the condensate fraction. The high $OD$ image was fit to a Thomas-Fermi (TF) profile combined with an independent Gaussian function. The pixels within a radius 10\% greater then TF radius were masked, and the remaining image was fit to a Bose-Einstein distribution to determine the number of thermal atoms $N_{th}$.  In images with sufficient signal, the low $OD$ image was fit to a combined TF--thermal profile, and the condensate number $N_{BEC}$ was determined from the TF component.  The condensate fraction was determined as $N_{\text{BEC}}/(N_{\text{th}}+N_{\text{BEC}})$ in this case. If the signal-to-noise ratio was too small to resolve the thermal component in the low $OD$ image, a fit to a simple TF profile was used determine the number of atoms $N_{\text{TF}}$ within the TF radius. The total number of atoms $N$ was determined in this case by adding the number of atoms outside the TF radius in the high $OD$ image. The condensate fraction was then $1-N_{\text{th}}/N$.  Entropy per particle was determined from the condensate fraction and total number using the semi-ideal model \cite{PhysRevA.58.2423}.

Peak fraction $f_0$ was determined from images of the gas released from the lattice using the procedure from Ref.~\cite{PhysRevA.85.061601}.  Several complications make interpreting fits to TOF images for lattice gases difficult. Unlike the trapped case, the functional form of the density distribution after TOF is unknown. Furthermore, atoms that appear outside of the condensate peaks are a combination of superfluid atoms (expelled from single-particle low-momentum states by strong interactions) and thermal atoms.  A final challenge is specific to our apparatus---there is no spherical symmetry in a lattice momentum distribution, and our imaging direction is not along a lattice direction. This makes even the non-interacting distribution difficult to fit (see Eq. 23 in Ref.~\cite{mckay:2009}).

\begin{figure}[h]
\includegraphics[width=0.8\textwidth]{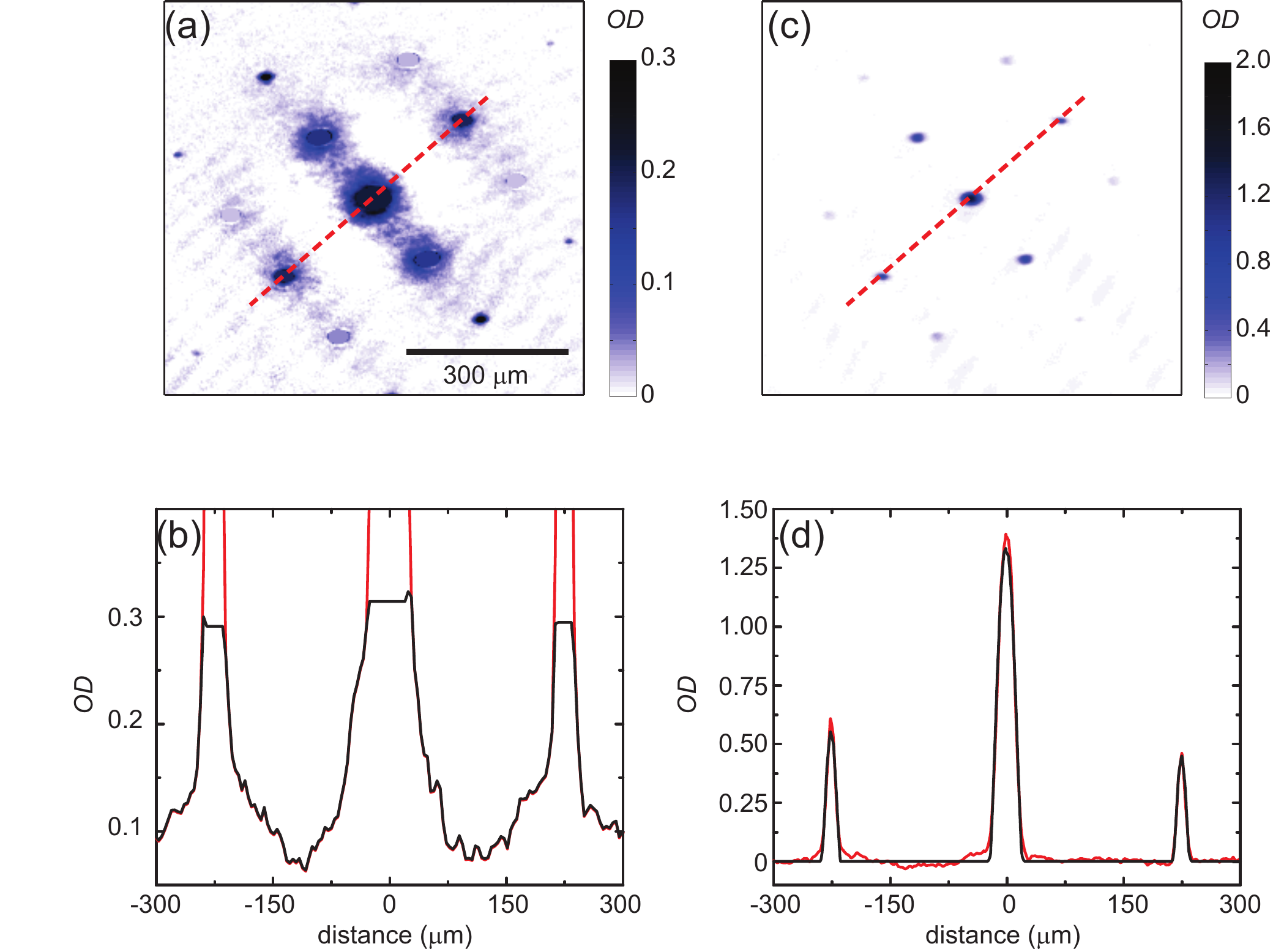}
\caption{(Color Online) Illustration of the fitting procedure used to determine peak fraction from a high $OD$ image (a and b) and from a low $OD$ image (c and d). Here we use an averaged image for $s=10$ and $S/Nk_B \approx 1.35$ (middle image in Figure \ref{fig:1}). For all these images, we have first subtracted off a Gaussian fit as described in the main text. In (a), we show the high OD image with the peaks masked and the non-peak distribution coincident with the peak assumed to be a uniform distribution. Note that we have masked the nine peaks that correspond to zero momentum and the first- and second-order diffraction peaks.  We ignore the higher-order peaks, which contain a negligible number of atoms. A slice along the red dashed line is displayed in (b). The black line is the non-peak distribution, and the red line is the complete image including the peaks. In (c), we show the low $OD$ image used to determined the peak number. A slice along the red dashed line is displayed in (d). The black line is the TF fit to the peaks and the red line is the data from the image. Note that there is a scale factor between the high and low OD images of approximately 5.75. \label{fig:2}}
\end{figure}

In the face of these difficulties, we have adopted a heuristic approach that is fast and independent of experiment details. In the high $OD$ image, we mask the peaks and fit the distribution to a single Gaussian. The number of atoms from this fit is $N_{\text{Gauss}}$. Next, the Gaussian fit is subtracted from the image and we fit each peak to a TF profile.  These fits are used to mask the peaks, and remaining signal is summed to determine the number of atoms $N_{\sum}$ outside of the peaks and not included in the Gaussian fit.  These steps capture all the non-peak atoms, except those that are part of the broader distribution and coincident with (or ``under") the peaks in the image.  Counting these atoms accurately would require knowledge of the density distribution after TOF, including the effects of interactions \cite{gerbier:2004,busch:2000,zawada:2008,PhysRevA.82.023618}.  Since analytical expressions for the TOF distribution are unknown, we make the simple assumption that the non-peak distribution that overlaps with the peak is uniform; see (a) and (b) of Fig. \ref{fig:2}.  We determine the average non-peak $OD$ coincident with the peak by averaging the $OD$ around the perimeter of the peak. The number of non-peak atoms $N_{\text{unpk}}$ coincident with the peak is estimated using a uniform distribution with this average $OD$. The total non-peak number of atoms is $N_{\text{nonpk}}=N_{\text{Gauss}}+N_{\sum}+N_{\text{unpk}}$.

We fit only the peaks of the low $OD$ image to determine the peak number $N_{pk}$ using a series of TF profiles (subtracting off a Gaussian if there is sufficient signal); see (c) and (d) of Figure \ref{fig:2}. The total number is then $N=N_{\text{pk}}+N_{\text{nonpk}}+N_{\sum}$, and the peak fraction is $f_0=1-N_{\text{nonpk}}/N$.  In the next section we demonstrate that $f_0$ is related to the condensate fraction $n_0$ by applying this fitting procedure to simulated TOF images produced using QMC calculations.

\section{Equilibrium Theory}

To assess whether equilibrium is achieved in the lattice, we determine peak fraction $f_0$ and entropy per particle $S/N$ for equilibrium gases using QMC simulations. We also compute condensate fraction $n_0$ in order to establish the relationship between condensate and peak fraction.  We match the parameters of the experiment as closely as possible. The particle number is kept constant within 5\% to $N = 200,000$ for optical lattice depths of $s = $ $6$,$8$,$10$ and $12$.  Furthermore, the confining potential is accounted for exactly in the QMC simulations.

QMC simulations are carried out using the stochastic series expansion method (SSE) \cite{sandvik:2002}. SSE is an exact method suffering from no biases or approximations. The condensate fraction $n_0$ is obtained directly from the single particle density matrix in adherence to its rigorous definition as the ratio of the occupation number of the macroscopically occupied single-particle mode to the total particle number. The general prescription used to calculate $n_0$ and the momentum distribution $n(\vec{k})$ that includes finite TOF effects has been discussed elsewhere \cite{Ray:2013}.

\begin{figure}[h]
\includegraphics[width=0.8\textwidth]{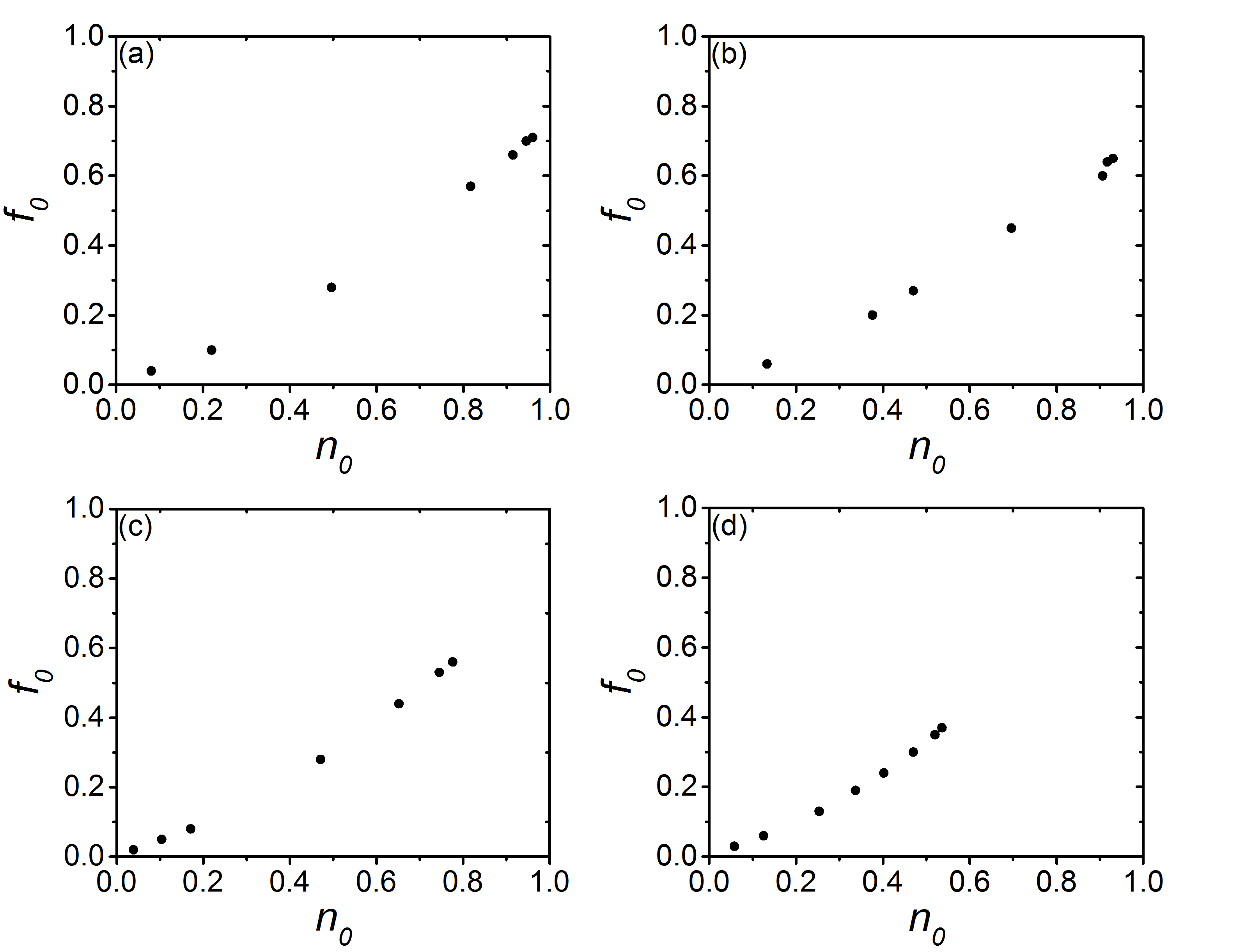}
\caption{Comparison of condensate $n_0$ and peak $f_0$ fraction obtained from QMC simulations for $s = 6,8,10,12$ (a-d, respectively). $n_0$ was obtained directly by diagonalizing the single particle density matrix.  $f_0$ was calculated by fitting QMC momentum distributions for a finite TOF of $20$ ms using experimental fitting procedures. The monotonic dependence is clear, but it is to be noted that the procedure cannot be used when $n_0 < 0.05$. The uncertainty in these values is smaller than 1\%.  The maximum $n_0$ and $f_0$ is smaller at higher $s$ because of quantum depletion.}
\label{fig:n0vsf0}
\end{figure}

We use the momentum distribution $n(\vec{k})$ from QMC simulations to extract the peak fraction using the fitting procedure applied to the experiment modified to work with a single image. Fig.~\ref{fig:n0vsf0} shows $n_0$ and $f_0$ computed at the same temperature for the $s$ sampled in the experiment. It is evident that $n_0$ and $f_0$ are monotonically related, and that peak fraction can be used to determine temperature in the experiment (assuming that equilibrium is achieved).  The relationship between $f_0$ and $n_0$ also does not depend strongly on lattice potential depth.  When $n_0 < 0.05$, the fitting procedure cannot identify the presence of condensate even with virtually noiseless QMC data.

Generally, $f_0$ is less than $n_0$.  This discrepancy between $f_0$ and $n_0$ arises from inaccurate accounting of the condensate and non-condensate distributions.  In part, this is due to the high-momentum tails of the condensate that extend into the broad Gaussian-like part of the momentum distribution, and that are mistakenly accounted for as non-condensate by the fitting procedure \cite{Ray:2013}. Apart from these systematic errors in the fitting protocol, there may also be additional interaction effects that arise during TOF that we cannot account for in the QMC simulations.

\begin{figure}[h]
\includegraphics[width=0.8\textwidth]{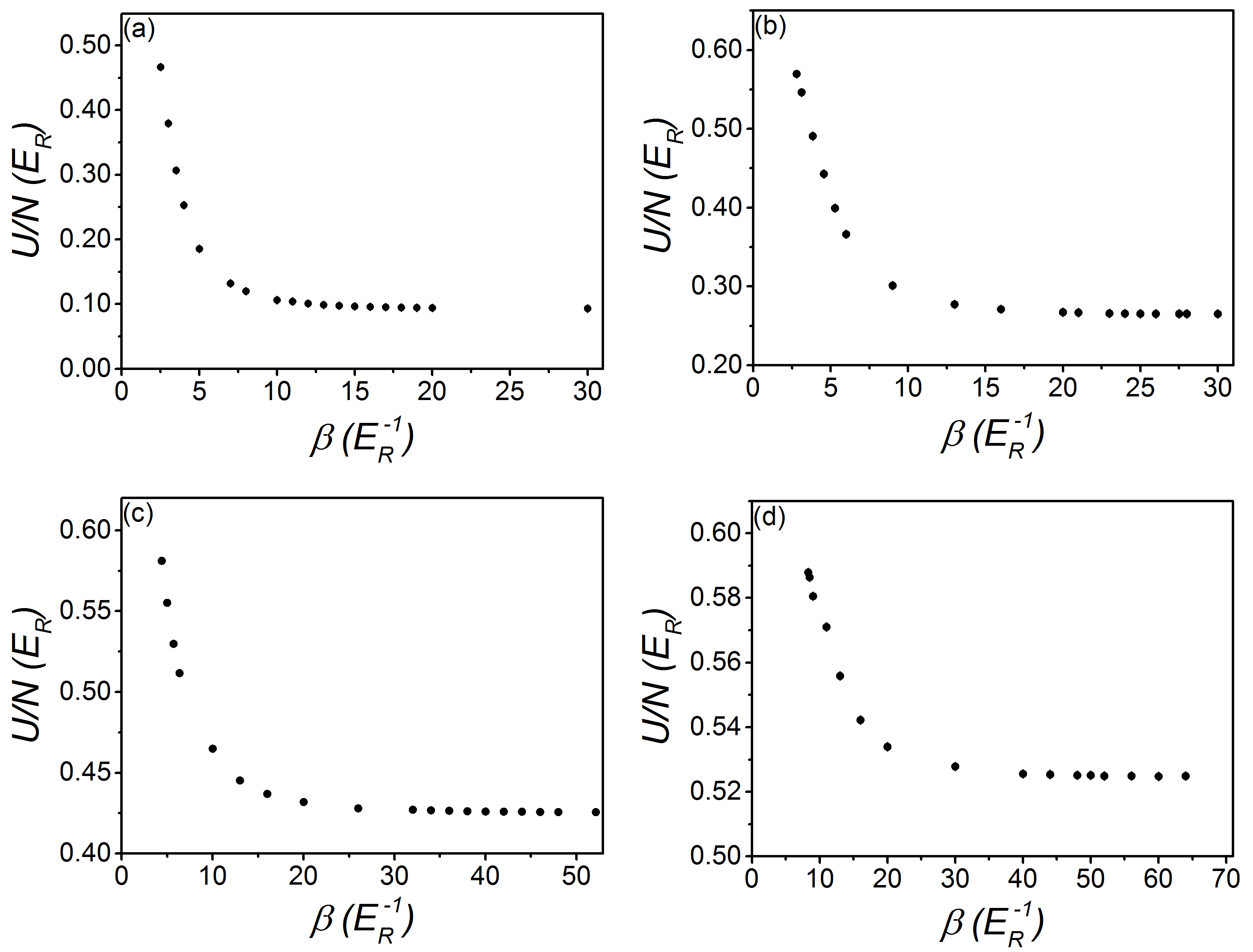}
\caption{The energy per particle for varying lattice depths $s = \{6,8,10,12\}$ (a-d, respectively) for a system of $N \sim 200,000$ particles in a trap.  These point were fitted using a procedure outlined in the text to obtain the entropy per particle that can be compared with experiment.  Temperatures only below the critical temperature for superfluidity are sampled.}
\label{fig:U_NvsiT}
\end{figure}

In order to compare with experiments that have access to the entropy per particle but not temperature, we also use QMC simulations to compute the temperature dependence of $S/N$. This is done by integrating the internal energy per particle ($u(\beta) \equiv U/N(\beta)$) shown in Fig.~\ref{fig:U_NvsiT}, obtained from QMC simulations, where $\beta=1/k_B T$.

Using the definition $u(\beta) = u_0 + f(\beta)$, where $u_0 \equiv lim_{\beta\rightarrow\infty} u(\beta)$ is the ground state energy and $f(\beta)$ is a monotonically decreasing function that characterizes the temperature dependence with $lim_{\beta\rightarrow\infty} f(\beta) \rightarrow 0$, we obtain
\begin{equation}
S/N(\beta) = \beta f(\beta)|_\infty^{\beta} + \int_\beta^{\infty} f(\beta) d\beta = \beta f(\beta) + \int_\beta^{\infty} f(\beta) d\beta
\end{equation}
Here the limit $lim_{\beta\rightarrow\infty} \beta f(\beta)$ can be taken provided $f(\beta)$ decays faster than linearly---a condition that is easily met since the data fit well to an exponential function with a reduced $\chi^2 \sim 1$.  Our strategy is to integrate $u$ using a fit to all sampled points. We fit the high $\beta$ tail to an exponential decay, since we cannot use a finite temperature method to access the true ground-state energy at $\beta\rightarrow\infty$. The high temperature points are fit using a cubic spline. Error bars for $S/N$ are computed using a resampling procedure. For each $u(\beta)$ that we have computed, we sample a gaussian of mean $u(\beta)$ and standard deviation $\sigma(u)$ to generate a sample $u_i(\beta)$. A collection of $u_i$ describe the vector $U_i \equiv \{u(\beta)\}$ that is fitted to obtain an estimate of $(S/N)_i$. This procedure is repeated multiple times to obtain an estimate of $\langle S/N(\beta)\rangle$ and $\sigma(S/N)$.

We compare the QMC results to a variety of approximate theories, in order to evaluate the accuracy of these less time consuming methods.  We consider  mean-field theories (MFT) viz., the Hartree-Fock (HF), Hartree-Fock-Bogoliubov-Popov (HFBP) and Gutzwiller approximations.  While these approaches have been used to analyze experiments as benchmarks for more sophisticated theories \cite{trotzky:2010,spielman:2008,white:2009,pethick:2008,rey:2003,pasienski:2010,gaul:2013}, their validity has not been comprehensively examined.  For calculations involving MFTs, the trap is accounted for using the local density approximation.

The HF and HFBP approaches rely on a perturbative expansion of the local field operator $\hat{\psi}(\vec{r}) = \psi_0(\vec{r}) + \delta\hat{\psi}(\vec{r})$ that accounts for the macroscopically occupied scalar condensed mode ($\psi_0(\vec{r})$) together with the fluctuations ($\delta\hat{\psi}(\vec{r})$). The former treats the kinetic energy terms of the BH Hamiltonian exactly, while treating the interactions via the HF approximation, which leads to the HF spectrum
\begin{equation}
 \varepsilon_{\text{HF}}(\vec{q})  =  \epsilon(\vec{q})-\mu+2U(\rho_0 + \rho_{\text{th}}).
\end{equation}  The HFBP approach allows for mixing of particle-like and hole-like excitations, which produces a dispersion
\begin{equation}
\varepsilon_{\text{HFBP}}(\vec{q})  =  \sqrt{(\varepsilon_{\text{HF}}(\vec{q}))^2-(U\rho_0)^2},
\label{ehfbp}
\end{equation}
where $\vec{q}$ is the quasi-momentum, $\rho_0$ is the condensate density, $\rho_{\text{th}}$ is the non-condensate density, $\mu$ is the chemical potential, and $\epsilon(\vec{q})=2t\left[3-cos(q_x d/\hbar)-cos(q_y d/\hbar)-cos(q_z d/\hbar)\right]$ is the tight-binding dispersion for single particles in the lowest energy band.

We obtain the non-condensed density via
\begin{equation}
\rho_{\text{th}} = \frac{1}{2} \sum_{\vec{q}\neq0} \left[ \frac{\varepsilon_{\text{HF}}(\vec{q})}{\varepsilon(\vec{q})}\coth(\beta\varepsilon(\vec{q})/2) - 1\right]
\label{mftsfeq}
\end{equation}
where $\varepsilon(\vec{q}) = \varepsilon_{\text{HF}}(\vec{q})$ ($\varepsilon_{\text{HFBP}}(\vec{q})$) for the HF (HFBP) calculation. Due to the appearance of $\rho_{\text{th}}$ on both sides of (\ref{mftsfeq}), it must be solved self-consistently, starting with a guess for $\rho_0$ and utilizing the relation $\mu = U(\rho_0 + 2\rho_{\text{th}})$. Using the spectrum determined using this procedure, we calculate relevant thermodynamic observables, such as the entropy
\begin{equation}
S  =  \sum_{\vec{q}\ne0} \left[ \frac{\beta \varepsilon(\vec{q})}{e^{\beta \varepsilon(\vec{q})}-1} - ln \left(1-e^{-\beta \varepsilon(\vec{q})}\right)\right],
\end{equation}
where $\beta=1/k_B T$.

The Gutzwiller approach utilizes a perturbative expansion of the kinetic term (thereby treating it approximately) while keeping the interaction term intact. We use $\hat{a}^{\dagger}_i\hat{a}_j \approx \langle \hat{a}^{\dagger}_i \rangle \hat{a}_j + \langle \hat{a}_i \rangle \hat{a}^{\dagger}_j - \langle \hat{a}^{\dagger}_i \rangle \langle \hat{a}_j \rangle$ to decouple the Hamiltonian, where $\hat{a}^{\dagger}_i$ and $\hat{a}^{\dagger}_j$ are particle creation operators on neighboring sites $i$ and $j$.  In this approximation, the Hamiltonian becomes independent of neighboring sites since $\langle a_j \rangle = \alpha$ is a complex number:
\begin{equation}
H \approx  -zt(\alpha \hat{a}^{\dagger} + \alpha^{*} \hat{a} - |\alpha|^2) + \frac{U}{2}\hat{n}(\hat{n}-1)-\mu \hat{n},
\end{equation}
where $z=6$ is the coordination number and $\hat{n}=\hat{a}^{\dagger}\hat{a}$.  Using a truncated occupation number basis, we  construct a matrix for $H$ that is  diagonalized utilizing the relation $\alpha = \sum_i \langle i | \hat{a} | i \rangle e^{-\beta E_i}$, needed to account for the finite temperature occupation of states. We solve this system iteratively, starting with a guess for $\alpha$ which is updated until convergence occurs. All thermodynamic quantities are computed using $\alpha$ at a specific $\beta$, $\mu$, and $U/t$.

In order to account for the parabolic trap used in experiments, we rely on the local density approximation (LDA) for MFT results. The LDA treats the slowly varying external confining potential as a local chemical potential shift.  Thermodynamic quantities are computed at different $\mu(r)$ (corresponding to a homogeneous system) and then integrated over all $r$ to determine observables for the trapped lattice gas. The LDA ignores correlations across the trap.

\begin{figure}[h!]
\includegraphics[width=0.8\textwidth]{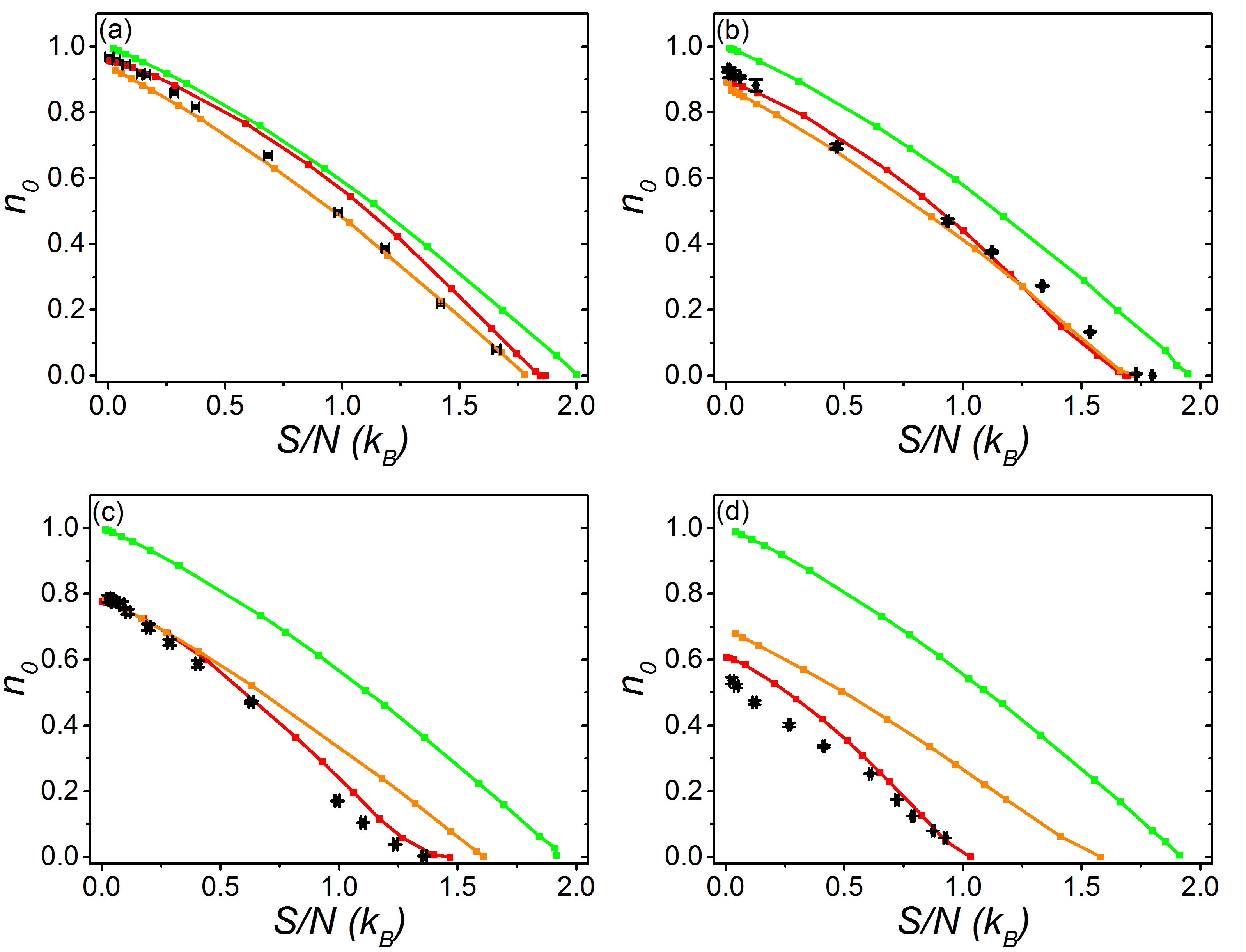}
\caption{(Color online) Comparison between different types of mean field theories against QMC results. Panels (a)-(d) correspond to $s=6$,$8$,$10$,$12$, respectively. The particle number was kept constant at $N \sim 200,000$. All other parameters were matched to experimental specifications presented in Fig. (\ref{fig:dataQMC_Expcomp}). Green curves correspond to HF, orange to HFBP, red to Gutzwiller, and black to QMC.}
\label{fig:datatheorycomp}
\end{figure}

We show predictions for condensate fraction $n_0$ at different entropies per particle and lattice potential depths in Fig.~\ref{fig:datatheorycomp} for QMC simulations and MFTs.  In each case, $T$ and $\mu$ are varied to match $S/N$ and $N$ as computed by the method of interest.  While all approaches show the same qualitative trend, it is clear that HF theory is, in general, not in good agreement with the QMC approach. This disagreement is expected because HF theory does not account for the mixing of particle and hole-like excitations that result from interactions. The comparisons of QMC against Gutzwiller and HFBP results suggest that for moderate interaction strengths (at $s = 6$ to $8$) these MFTs work generally well. For stronger interactions (at $s = 10$), HFBP calculations fail to capture finite temperature effects, whereas the Gutzwiller approximation is able to capture qualitative features. At the highest lattice potential depths we sample ($s = 12$), the Gutzwiller method can capture the high temperature regime, but cannot accurately compute the correlations at low temperature, as is evidenced by the 15-25\%  error in $n_0$ for $S/N<0.5$~$k_B$.

\begin{figure}[h!]
\includegraphics[width=0.8\textwidth]{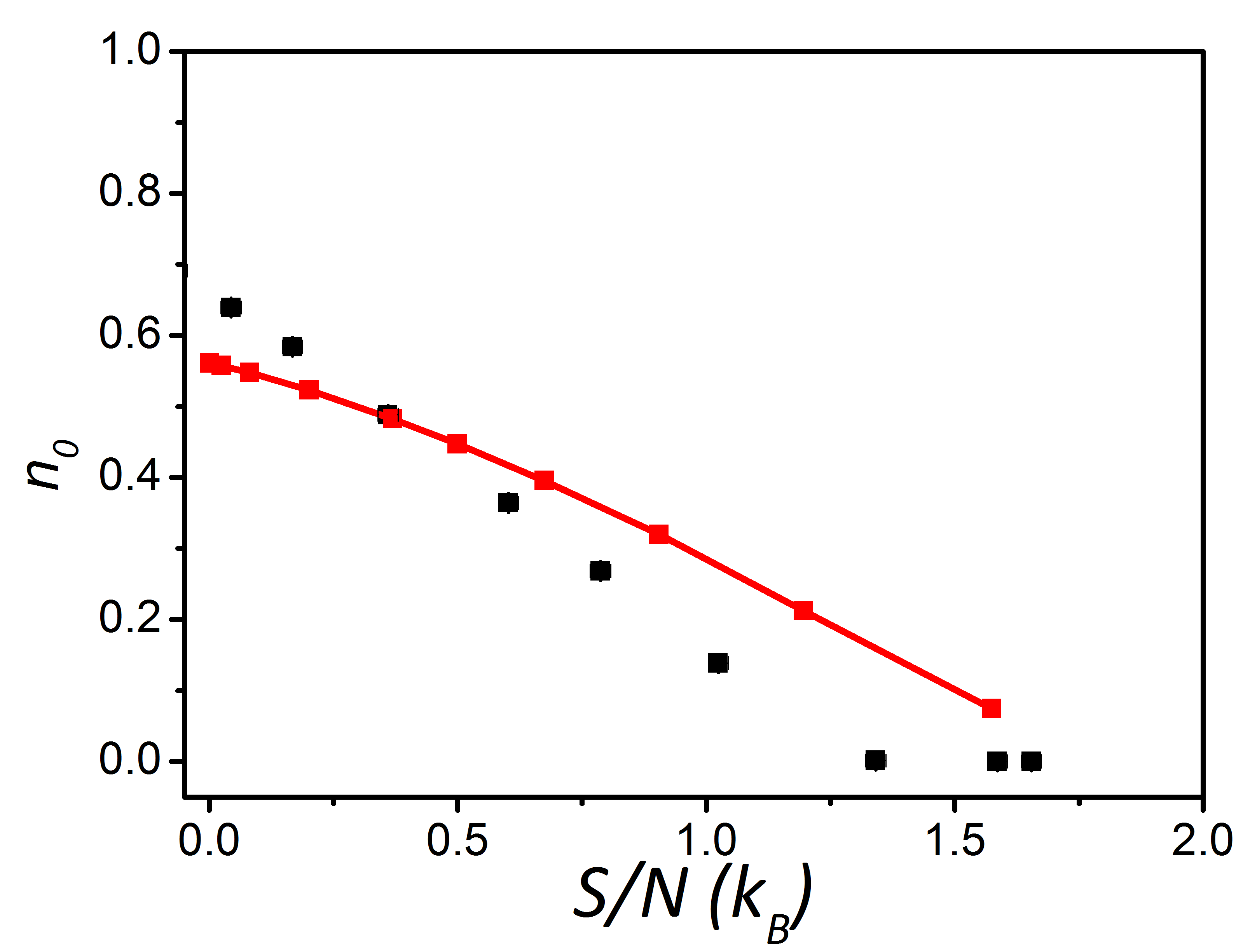}
\caption{(Color online) Comparison between Gutzwiller (red) and QMC (black) results for a strongly interacting system ($s = 13.6$) at low filling.  The central density for these results is approximately one particle per site with $\omega=2\pi\times55.6$~$Hz$ and $N \sim 60,000$.  The line is a guide to the eye.
}
\label{fig:datatheorycomp2}
\end{figure}

While the validity of the Gutzwiller approximation over a wide range of interaction strengths and temperatures is suggestive, this is only likely true for the high filling limit (a central density of approximately 3 particles per site) considered in this work. QMC simulations for low filling systems show that the differences with the Gutzwiller approximation can be quite large, as evident for the results for a central density of one particle per site shown in Fig.~\ref{fig:datatheorycomp2}.  The discrepancy apparent in Fig.~6 may also arise from quantum critical behavior, as the lattice potential depth is set close to the MI--SF phase transition.

We conclude that the HF and HFBP approximations should not be used to compute observables such as $n_0$ and thermodynamic variables like $S/N$.  While the Gutzwiller approximation can provide quantitatively accurate results at high filling and for moderate lattice potential depths, it fails at low $S/N$ and for strong interactions and for low densities.  We therefore use only QMC simulation results to compute $S/N$ and $f_0$ for comparisons with the experimental data.

\section{Results and Discussion}

\begin{figure}[h!]
\includegraphics[width=0.8\textwidth]{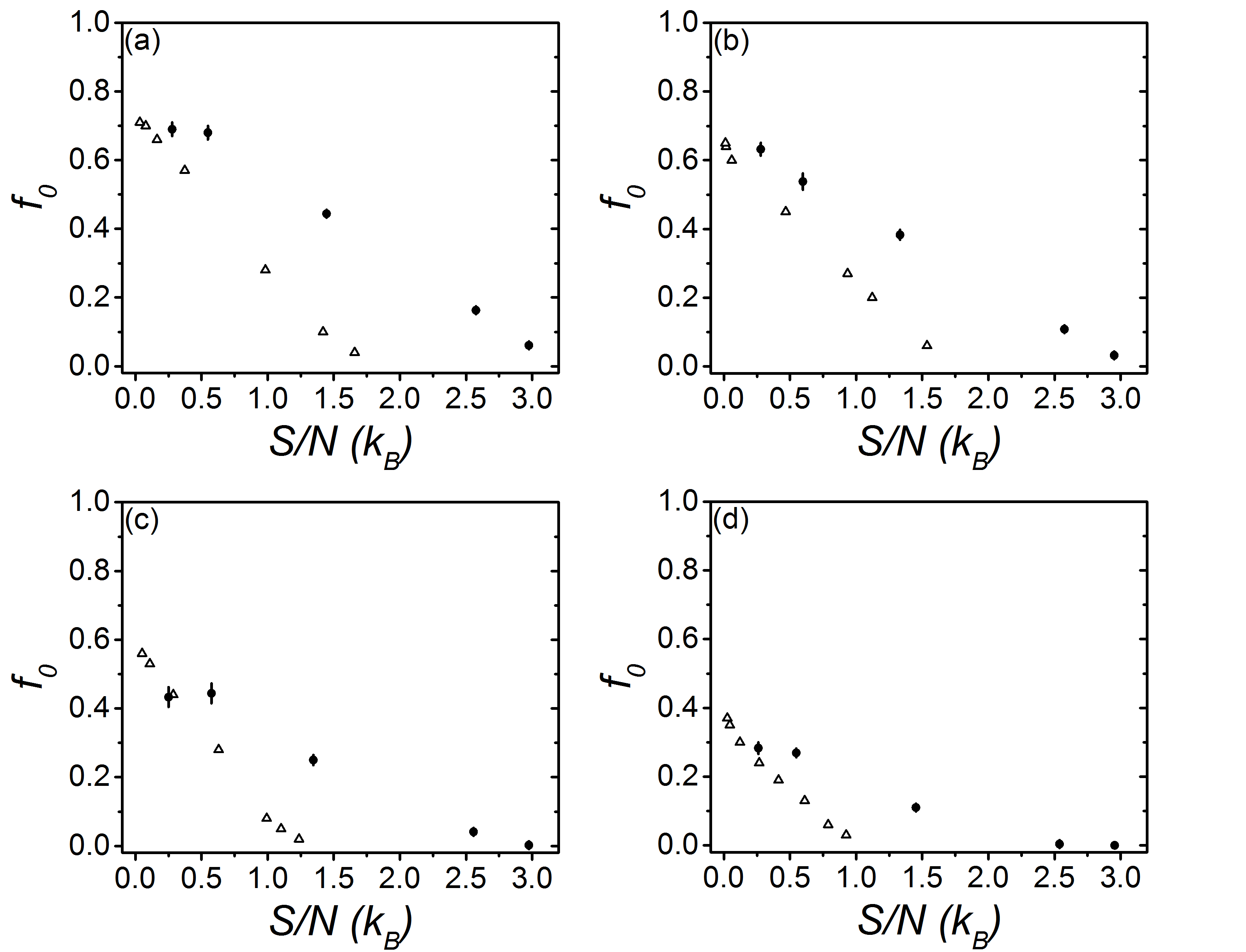}
\caption{Experimentally determined peak fraction at each lattice depth as a function of the initial entropy per particle in the harmonic trap (circles). Corresponding peak fraction from QMC simulations determined by fitting the momentum distribution using the same procedure used in experiments in the lattice are also shown (triangles).
The number of atoms at each experimental point (from lowest to highest entropy) is
(a) $s=6$, $N=\{2.38\pm0.03,2.08\pm0.06,1.64\pm0.03,2.04\pm0.07,2.06\pm0.09\}\times10^5$,
(b) $s=8$, $N=\{2.36\pm0.06,2.05\pm0.11,1.66\pm0.07,2.02\pm0.09,1.73\pm0.05\}\times10^5$,
(c) $s=10$, $N=\{2.41\pm0.08,2.15\pm0.11,1.68\pm0.09,2.03\pm0.16,1.85\pm0.09\}\times10^5$ and, for
(d) $s=12$, $N=\{2.55\pm0.12,1.98\pm0.10,1.65\pm0.04,2.06\pm0.11,1.74\pm0.11\}\times10^5$.
QMC curves are constrained to a constant number, which is the mean of the number at each lattice depth ($N \sim 200,000$). The tunneling, interaction energies and trap frequencies for $s=\{6,8,10,12\}$ respectively are $t=\{1.17,0.711,0.443,0.283\}\times10^{-31}$J, $U=\{3.92,5.23,6.48,7.67\}\times10^{-31}$J and $\omega=2\pi\times\{51.27,55.48,59.4,63.07\}$Hz.  The error bars in the experimental data show the standard error of the mean for the average across 7 measurements.  The uncertainty in the QMC results are too small to be visible.}
\label{fig:dataQMC_Expcomp}
\end{figure}

To assess thermal equilibrium in the experiment, we show a comparison between the experimentally measured $f_0$ (determined from images shown in Fig. 1 and the fitting procedure described in Sec. I) and the equilibrium QMC prediction at the same $S/N$ in Fig.~\ref{fig:dataQMC_Expcomp}.  A strong discrepancy is evident: there is alarming degree of disagreement between the QMC results and the experiment at high temperatures (i.e., high $S/N$). The experimentally observed peak fraction (and therefore condensate fraction) are systematically high compared with the QMC prediction. What is more troubling is that the condensate survives beyond the transition temperature, which is marked by $f_0$ (and $n_0$) vanishing in the QMC predictions.

We do not believe that this discrepancy can be explained by systematic errors in the analysis of the experimental data.  One source of error might be that binary collisions between atoms in different peaks eject condensate atoms during TOF \cite{greiner:2001,band:2000}.  While this effect may play a role in suppressing peak fraction for the lowest entropies at $s=6$, its impact should be reduced at higher lattice potential depths and temperatures as the condensate density decreases.  Furthermore, this effect would lead to lower, not higher, peak fractions than the equilibrium QMC prediction.

Another possibility is that we erroneously identify peaks in the TOF image.  It has been previously discussed that sharp peaks in momentum distributions may not be uniquely associated with a condensate \cite{kato:2008,diener:2007}.  However, calculations focusing on trapped systems for realistic experimental conditions and experiments have demonstrated that the visibility, which is a quantitative measure of peak sharpness, can distinguish between condensed and non-condensed states \cite{pollet:2008,gerbier:2007,trotzky:2010}.  The features we observe at high $S/N$ are narrow and quantitatively similar in extent to the peaks observed at low temperatures.  The sharpness of the peaks we measure at high $S/N$ is most apparent in the $s=6$ and $s=8$ images shown in Fig. 1.  This bimodal nature of the TOF images is associated with the existence of a condensate \cite{lin:2008}, and thus it is highly unlikely that we are systematically finding a non-zero peak fraction in images where a condensate is absent.  Furthermore, we do not observe peaks in the TOF distribution determined from QMC above $T_c$.

A further potential source of systematic error is heating by the lattice light \cite{mckay:2011}, which would have the largest impact at low entropy. However, we find that the QMC equilibrium predictions and the experimental data agree well in this regime.  For higher entropies, heating should reduce peak fraction below the QMC prediction, which is opposite to the observed behavior.

We conclude that the measured peak fraction is systematically higher than that allowed in equilibrium by the second law of thermodynamics---the measured $f_0$ implies that the gas is at lower $S/N$ after the lattice is turned on.  Furthermore, a condensate appears present at temperatures exceeding the critical temperature for superfluidity in the lattice.  Condensate and peak fraction should always decrease in equilibrium as the lattice is turned on to higher $s$ because of quantum depletion resulting from strong inter-particle interactions.

The experimental observations imply that equilibrium is not established and the condensate is metastable during the lattice turn on.  A natural question is whether the lattice is turned on too quickly for thermalization to take place.  One limitation is non-adabaticity associated with changes in the size of the gas and mass flow \cite{natu:2011}.  This problem is primarily associated with the Mott-insulator regime, which we do not sample here.  Furthermore, we estimate that overall rms radius of the gas changes by less than 20\% at $s=6$ and $S/N\approx1.5$~$k_B$ when the lattice is turned on, which is not likely to cause the large discrepancy evident in Fig. 7 for these parameters.

\begin{figure}[h!]
\includegraphics[width=0.8\textwidth]{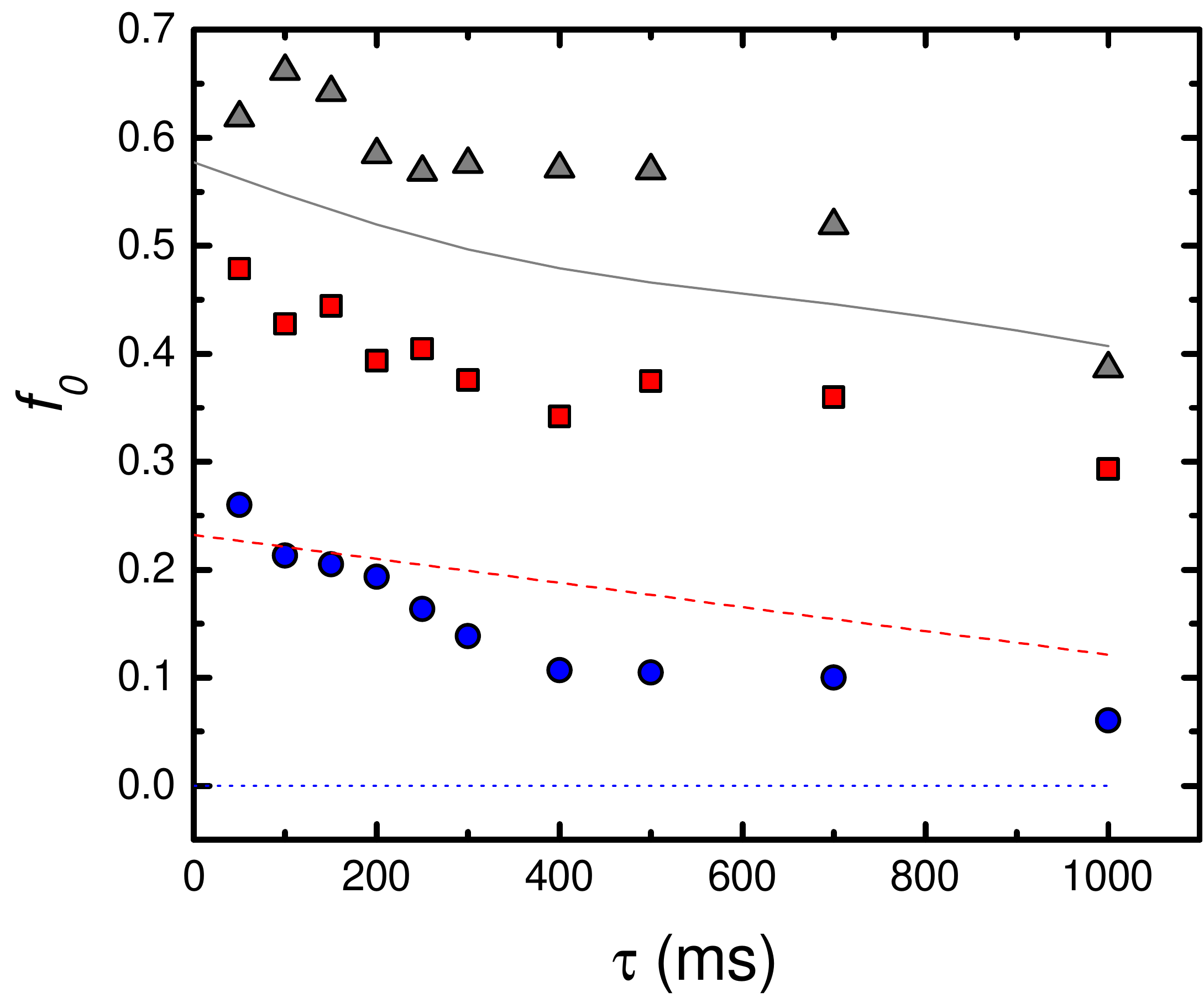}
\caption{(Color online) Peak fraction for different lattice turn on times $\tau$ at $s=6$.  Data and theoretical predictions are shown for $S/N$=0.44 (triangles, solid), 1.2 (squares, dashed), and 2.2 (circles, dotted) $k_B$.  The lines are the equilibrium $f_0$ predicted by QMC simulations taking into account heating from the lattice light.  At the highest $S/N$, QMC predicts that a condensate should be absent in the lattice.}
\label{fig:f0vsholdTime}
\end{figure}

Another issue is how the lattice turn-on time compares with natural timescales in the Hamiltonian.  The 100 ms lattice turn-on time is slow compared with the Hubbard timescales $h/t$ and $h/U$, which vary from 6--24 ms and 2--1 ms for $s=6$--12.  The many-particle timescale for thermalization may be longer than these single- and two-particle times, however.  We examine this possibility at $s=6$ by turning on the lattice over times up to 1~s for different $S/N$ in the trap.  A shown in Fig.~\ref{fig:f0vsholdTime}, we include data for low $S/N$ where the agreement with QMC predictions is good, and high $S/N$ where the experimental peak fraction is non-zero above the critical temperature.

In all cases, no adiabatic timescale is apparent---the peak fraction continuously decreases with the lattice turn-on time. This behavior is characteristic for all the $s$ sampled in this work.  At low and intermediate $S/N$, the decrease in peak fraction is consistent with the heating from the lattice light.  We compute the predicted equilibrium peak fraction including heating for different lattice turn on times $\tau$ using QMC results for $u$.  The rate of increase in $u$ from the lattice light and the corresponding increase in $S/N$ and decrease in $f_0$ are determined using the results from Ref.~\cite{mckay:2011}. We conclude that the adiabatic timescale is comparable to or longer than the timescale for heating, and thus discriminating between the two processes is not possible.

Ultimately, inter-particle interactions via binary collisions are the dominant mechanism by which equilibrium is established and condensate fraction decreases as the lattice potential is turned on.  Despite extraordinary advances in methods such as QMC, computing exact dynamics in higher than one dimension for more than a few tens of particles and short times remains infeasible for strongly interacting systems such as the BH model.  In the next section, we use a hybrid of equilibrium QMC simulations and approximate theory to compute the timescale associated with the simplest mechanism---Landau damping---for relaxation of $n_0$.

\section{Landau Damping Theory for Relaxation of the Condensate}

In this section, we outline a model for relaxation of the condensate atom number via Landau damping. In a closed system, the condensate number can change by collisions with quasi-particles, whose number is not conserved. The dominant collision processes involve two-body scattering, where a condensate atom collides with a quasi-particle to produce two new quasi-particles. In equilibrium, the rates for this process and its reverse (whereby two quasi-particles collide to produce a condensate atom and a quasi-particle) are identical but opposite in sign, and the condensate number is conserved. However out of equilibrium, these processes do not cancel, and induce changes in the condensate atom number.

In general, the theoretical problem of how the condensate interacts with the thermal gas is an extremely challenging one, but an understanding of the precise mechanisms for equilibration is paramount to current experiments in optical lattices. Briefly, a condensate can relax by colliding with thermal atoms in two ways: at the mean-field level, the condensate mean-field introduces fluctuations in the thermal cloud which in turn lead to a damping of the condensate. This process is termed Landau damping, which is what we calculate. There is a second process where the condensate atoms exchange energy with the thermal atoms via collisions \cite{griffin:1999,griffin:2001}. Although we do not calculate this process here, it is roughly of the same order as the Landau damping mechanism we compute \cite{griffin:2001}. A third process is called Belieav damping, whereby a non-condensed atom collides with a condensate atom and decays into two non-condensed atoms with lower energy. This mechanism is dominant at very low temperatures and its rate decreases rapidly with temperature.  Calculating the Landau damping rate thus suffices to estimate the timescales for equilibration in lattices.

\begin{figure}[h!]
\includegraphics[width=0.8\textwidth]{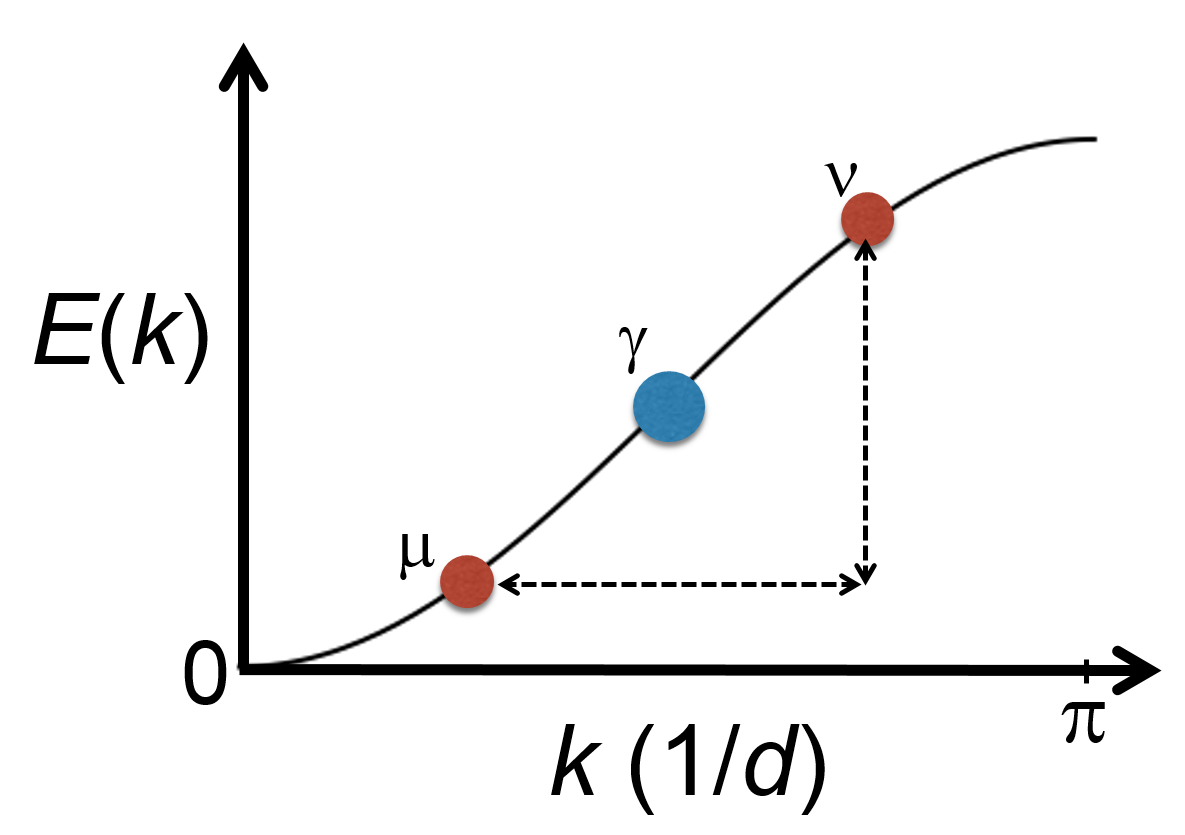}
\caption{(Color online) Schematic representation of Landau damping in one dimension.  The dispersion $E(k)$ is shown using a solid line.  We consider the process whereby a quasiparticle ($\gamma$) can decay by coupling to a resonant transition between two quasiparticles ($\nu$ and $\mu$) mediated by the underlying condensate.  By conservation of total particle number, this reduces the overall condensate density.}
\label{fig:LD}
\end{figure}

Our treatment of Landau damping directly follows that of Tsuchiya and Griffin \cite{griffin:2005}, who computed the damping rate for a homogeneous lattice Bose gas. Physically, this rate corresponds to the annihilation of a Bogoliubov quasi-particle and a condensate atom to produce two Bogoliubov quasi-particles (Fig.~\ref{fig:LD}). By conservation of total particle number, integrating the damping rate over all momenta yields the rate at which the condensate fraction changes via Landau damping. For a homogeneous gas, the Landau damping rate for a quasi-particle at momentum $\vec{q}$ is given by \cite{griffin:2005}:
\begin{widetext}
\begin{eqnarray}\label{ldamping}
\Gamma^{\vec{q}}_{L} & = & \pi n_{0} \sum_{\vec{p}, \vec{k}}|M_{\vec{q},\vec{k}; \vec{p}}|^{2}\delta\left\{\varepsilon(\vec{p}) - \left[\varepsilon(\vec{p}) -\varepsilon(\vec{k})\right]\right\}[f^{0}(\varepsilon(\vec{k}))-f^{0}(\varepsilon(\vec{k}))], \\\nonumber
M_{\vec{q},\vec{k}; \vec{p}} & = & \frac{2U}{\sqrt{L^{d}}}\sum_{\vec{G}}~\Big[(u_{\vec{q}}u_{\vec{p}} + v_{\vec{q}}v_{\vec{p}} - v_{\vec{q}}u_{\vec{p}})u_{\vec{k}} - (u_{\vec{q}}u_{\vec{p}} + v_{\vec{q}}v_{\vec{p}} - u_{\vec{q}}v_{\vec{p}})v_{\vec{k}}\Big]\delta(\vec{q} + \vec{k} - \vec{p}-\vec{G}),
\end{eqnarray}
\end{widetext}
where the $\vec{G}$ are reciprocal lattice vectors (which incorporate Umklapp scattering), $f^{0}(x) = (\exp(\beta x)-1)^{-1}$ is the Bose distribution at temperature $k_{B}T = 1/\beta$, and $\varepsilon(\vec{k})$ are the usual Bogoliubov quasi-particle dispersions in a lattice of the form in Eq.~\ref{ehfbp}. The Bogoliubov coefficients $u_{\vec{k}}$ an $v_{\vec{k}}$ are given by $u_{\vec{k}} = \sqrt{\frac{1}{2}(\frac{\varepsilon_{\vec{k}} + Un_{0}}{\varepsilon(\vec{k})} + 1)}$ and $v_{\vec{k}}= \sqrt{\frac{1}{2}(\frac{\varepsilon_{\vec{k}} + Un_{0}}{\varepsilon(\vec{k})} - 1)}$. The delta functions enforce energy and momentum conservation.

The detailed derivation of this result can be found in Ref. \cite{griffin:2005} and is not repeated here. Note that a key assumption in this theory is that thermal (i.e., non-condensed) gas is in local equilibrium, which is enforced on short timescales by collisions between thermal atoms. We emphasize here that we are not modeling the full far-from-equilibrium problem of relaxation of condensed and non-condensed atoms following a lattice ramp. Rather, we compute the timescales for relaxation when the condensate density departs slightly from its equilibrium value at a given lattice depth. This near-equilibrium computation is the first step in gaining a systematic understanding of relaxation times in generic non-equilibrium situations.

We numerically compute the damping rate for all momenta $\vec{q}$ ranging from $-\pi/d \leq q_{i} \leq \pi/d$, where $i = \{x, y, z\}$, and $d$ is the lattice constant. We discretize momenta in steps of $\pi/L$, where $L$ is the system size, defined as where the total density vanishes. The finite system size also sets an infrared cutoff, $q_{\text{min}} = \pi/L$, which renders the one-dimensional integrals convergent. We have checked numerically that this cutoff is small enough that changing $q_{\text{min}}$ does not introduce significant errors. The total damping rate is given by $\Gamma_{L} =\sum_{\vec{q}}\Gamma(\vec{q})/\Omega$, where $\Omega$ is the system size. Using a simple rate equation $d\delta \rho_{0}/dt = - \Gamma_{L}\delta \rho_{0}$, where $\delta \rho_{0}(t) = \rho_{0}(t) - \rho^{\text{eq}}_{0}$, we obtain an estimate for the relaxation time $\tau_{LD}=1/\Gamma_{L}$ of the condensate.

To compute the damping rate, we require three quantities: the condensate density $\rho_{0}$, lattice depth $s$, and the temperature $T$. For a given lattice depth $s$ and temperature $T$, we use $\rho_0$ from exact QMC simulations discussed in Sec. III to compute the different Landau damping rate. Using the relation between temperature and entropy per particle, we obtain the damping rate as a function of entropy per particle for different lattice depths.

Our derivation of the damping rate assumes that all of the condensate atoms are at $\vec{q} = 0$. While this is true in a weakly interacting homogeneous system at low temperature, the presence of a trap and strong correlations modifies this picture significantly, as momentum is no longer a good quantum number. The condensate therefore develops a spread in momentum, which we have computed numerically using QMC for a given temperature and lattice depth.

To incorporate trap and interaction effects into our damping calculations, we produce the Landau damping rate in two cases: 1) we assume all the condensate atoms are at $\vec{q} = 0$. This yields an \textit{upper bound} on the damping rate. 2) we obtain a \textit{lower bound} on the damping rate by assuming that the total condensate number is equal to only the contribution to the momentum distribution at $\vec{q} = 0$. The experimentally observed damping rate must be between these two bounds.

\begin{figure}[h!]
\includegraphics[width=0.8\textwidth]{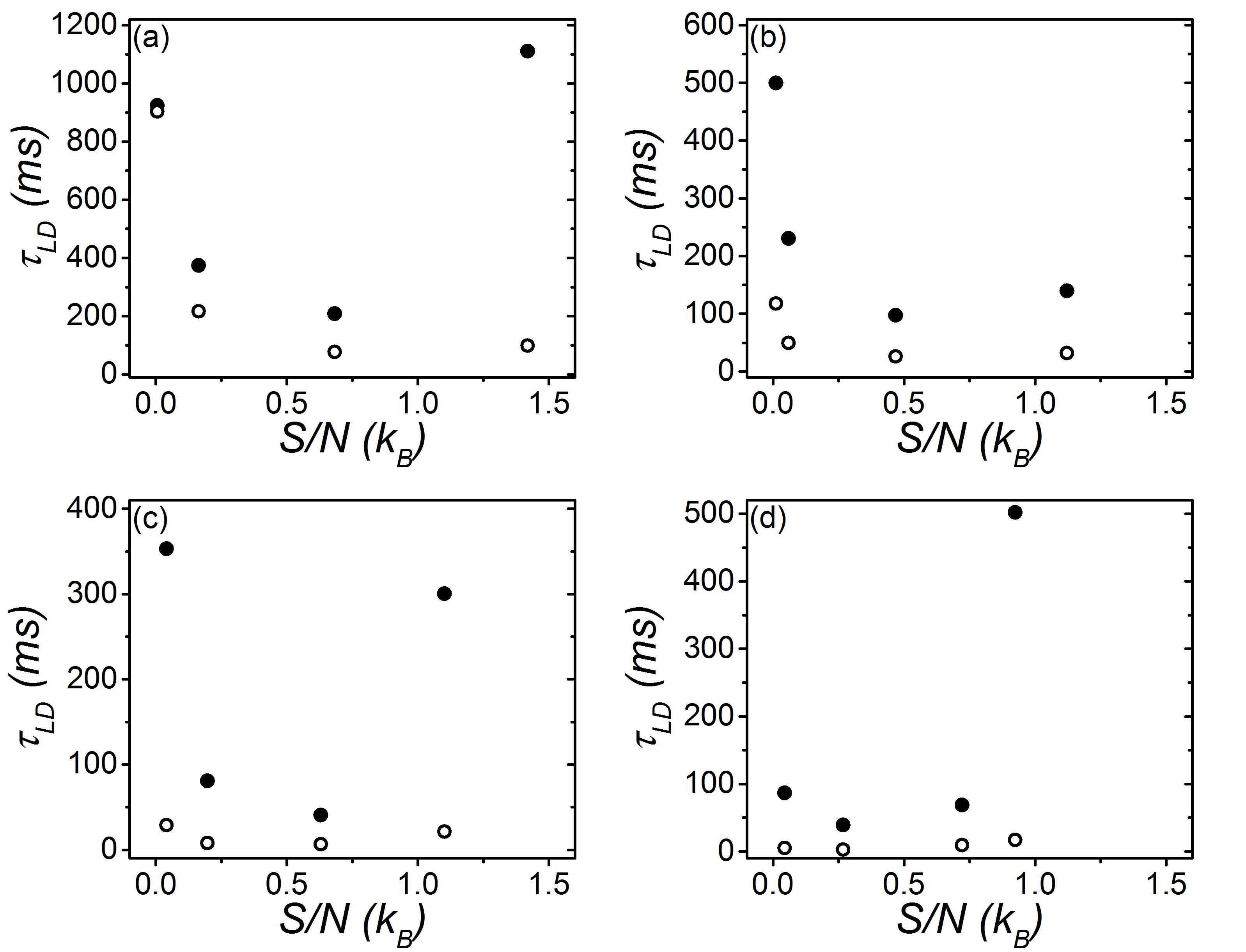}
\caption{Estimates of the lower (empty circles) and upper bound (filled circles) of the decay time associate with Landau damping vs. entropy per particle for $s = 6,8,10,12$ (a-d). }
\label{fig:LDvsS}
\end{figure}

The results of our calculations presented in Fig. \ref{fig:LDvsS} paint a fascinating picture. At low entropy, the lower bound on the Landau damping time ranges from $200$~ms to $10$~ms from the lowest to highest lattice depth $s = 6E_{R}$ to $s = 12$, while the upper bound varies roughly from $1$~s to $100$~ms. At the highest $S/N$, the Landau damping time is at least 20 times larger than the Hubbard tunneling time $h/t$.  The decrease in the damping time for higher lattice potential depth is consistent with the increase in interaction strength. For $s = 6$, where the system can be treated as weakly interacting, the two bounds coincide at low entropy because the condensate largely resides at zero momentum.  Interactions induce a significant broadening of the condensate momentum distribution, which leads to a large difference between the upper and lower bounds at higher $s$.  This is particularly stark at high entropies per particle, because the condensate is almost entirely absent. Note that for $n_{0} = 0$, $\Gamma_{L} \rightarrow \infty$. Strictly speaking, this theory is not accurate when the condensate number is small, either because of non-zero temperature or strong interactions.  At high temperatures, interactions between non-condensate atoms play an important role that is not captured by our approach.  For the strongly interacting case, quantum depletion reduces $n_0$ and the Bogoliubov spectrum, which is the underlying assumption in this theory, is not valid.  Therefore the upper bounds in the damping time at $S/Nk_{B} = 1$ for $s = 10$ and $12$ may not be correct.

Comparing the predicted Landau damping times to the 100~ms lattice turn-on used for the data in Fig. \ref{fig:f0vsholdTime} is complicated by the lack of equilibrium in the lattice.  Generally, the damping time exceeds 100~ms at the highest $S/N$, where the discrepancy between the QMC and experimental $f_0$ is greatest.  This implies that the lattice turn-on time used for Fig. \ref{fig:dataQMC_Expcomp} was non-adiabatic, since times much longer than $\tau_{LD}$ are required to achieve equilibrium in the lattice.  The predicted Landau damping time is also long at low $S/N$, but $n_0$ changes less in this regime as the lattice is turned on, and thus the impact of non-adiabatic behavior is less apparent.  Diabaticity with respect to Landau damping at high $S/N$ explains the experimental observation that the peak and condensate fraction in the lattice is higher than the equilibrium prediction, since the condensate does not have sufficient time to decay as the lattice is turned on.

The upper bound on the damping time in combination with the heating rate from the lattice light implies that adiabaticity may be impossible to achieve under certain conditions.  At $s=6$, for example, the upper bound on the Landau damping time at $S/N>1$ is close to 1 s.  This time is longer than the heating timescale observed for the data shown in Fig.~\ref{fig:f0vsholdTime}.  Differentiating between heating and thermalization becomes more challenging at higher $s$, since the characteristic energy scale $t$ drops rapidly, but the heating rate from the lattice light grows linearly with $s$.

\section{Conclusions}

In conclusion, we have experimentally and theoretically investigated peak fraction in the superfluid regime of a cubic lattice for varied entropy per particle and lattice potential depth.  Qualitatively, the experimentally observed peak fraction behaved as theoretically expected: lower initial entropies led to higher peak fractions, and higher lattice depths (lower $t/U$) resulted in lower peak fractions. However, we did not find quantitative agreement with exact QMC simulations. We observe that the condensate is metastable, and that non-zero peak fraction persists above the critical temperature for superfluidity for times much longer than the Hubbard timescales.  We have investigated the timescales for relaxation of the condensate by calculating the Landau damping time in the lattice. Our calculations indicate that the relaxation time is comparable or greater than typical lattice turn-on times.   The Landau relaxation time is also longer than the timescale associated with heating from the lattice light, which may make adiabaticity impossible to achieve.

Going forward, our theory-experiment comparison points to the urgent need to carefully understand timescales for dynamics in strongly correlated optical lattice experiments.  Furthermore, using this model as a starting point, one can envision branching out to situations where competing theories disagree, break down or do not even exist. A canonical example is that of strongly correlated systems far-from-equilibrium, where the very paradigms for how to think about these systems are only now being developed. In addition to being of fundamental importance, questions of non-equilibrium dynamics serve a practical purpose in cold atom experiments, which are largely isolated from the environment. Understanding the mechanisms for driving these systems to equilibrium is thus essential in order to understand timescales for maintaining adiabaticity.

\begin{acknowledgments}
The authors acknowledge funding from the NSF, ARO, and DARPA OLE program.  Computer simulations and data analysis were made possible by workstations supported by the DARPA-OLE program, as well as XSEDE resources at TACC (Texas). David McKay and Ushnish Ray contributed equally to this work.
\end{acknowledgments}

\bibliography{lattice_snapoff}

\end{document}